# FedDA-TSformer: Federated Domain Adaptation with Vision TimeSformer for Left Ventricle Segmentation on Gated Myocardial Perfusion SPECT Image

Yehong Huang[1], Chen Zhao[1*], Rochak Dhakal[2], Min Zhao[3], Guang-Uei Hung[4], Zhixin Jiang[5], Weihua Zhou[2,6]

1. Department of Computer Science, Kennesaw State University, Marietta, GA, USA

2. Department of Applied Computing, Michigan Technological University, Houghton, MI, USA

3. Department of Nuclear Medicine, Xiangya Hospital, Central South University, Changsha, China

4. Department of Nuclear Medicine, Chang Bing Show Chwan Memorial Hospital, Changhua, Taiwan

5. Department of Cardiology, The First Affiliated Hospital of Nanjing Medical University (Jiangsu Provincial Hospital), Nanjing, China

6. Center for Biocomputing and Digital Health, Institute of Computing and Cyber-systems, and Health Research Institute, Michigan Technological University, Houghton, MI, USA

\* Correspondence:

Chen Zhao, Ph.D.

Kennesaw State University

Email address: czhao4@kennesaw.edu

Mailing address: 680 Arntson Dr, Atrium BLDG, Marietta, GA 30060, USA

**Abstract**

**Background and Purpose:** Functional assessment of the left ventricle using gated myocardial perfusion (MPS) single-photon emission computed tomography relies on the precise extraction of the left ventricular contours while simultaneously ensuring the security of patient data.

**Methods:** In this paper, we introduce the integration of **Fed**erated **D**omain **A**daptation with **T**ime**S**former, named "FedDA-TSformer" for left ventricle segmentation using MPS. FedDA-TSformer captures spatial and temporal features in gated MPS images, leveraging spatial attention, temporal attention, and federated learning for improved domain adaptation while ensuring patient data security. In detail, we employed Divide-Space-Time-Attention mechanism to extract spatio-temporal correlations from the multi-centered MPS datasets, ensuring that predictions are spatio-temporally consistent. To achieve domain adaptation, we align the model output on MPS from three different centers using local maximum mean discrepancy (LMMD) loss. This approach effectively addresses the dual requirements of federated learning and domain adaptation, enhancing the model's performance during training with multi-site datasets while ensuring the protection of data from different hospitals.

**Results:** Our FedDA-TSformer was trained and evaluated using MPS datasets collected from three hospitals, comprising a total of 150 subjects. Each subject's cardiac cycle was divided into eight gates. The model achieved Dice Similarity Coefficients (DSC) of 0.842 and 0.907 for left ventricular (LV) endocardium and epicardium segmentation, respectively.

**Conclusion:** Our proposed FedDA-TSformer model addresses the challenge of multi-center generalization, ensures patient data privacy protection, and demonstrates effectiveness in left ventricular (LV) segmentation.

## 1. Introduction

Coronary artery disease (CAD) remains one of the leading causes of death and disability globally, significantly impacting healthcare systems worldwide [1]. Gated myocardial perfusion single-photon (SPECT) emission computed tomography (SPECT) has emerged as a valuable non-invasive diagnostic technique for evaluating CAD due to its favorable balance of accuracy and cost-effectiveness [2]. This imaging modality provides critical insights into left ventricular (LV) function, myocardial perfusion, and blood flow. Central to its utility is the precise delineation of cardiac structures such as the endocardium, myocardium, and epicardium, which enables accurate quantitative assessments of LV function [3]. However, manual segmentation is not only labor-intensive but also subject to variability among observers, both within and between individuals, leading to inconsistent results. Additionally, segmentation accuracy may be affected by challenges such as extracardiac activity and reduced tracer uptake, which can hinder the reliability of functional assessments.

With the advancement of deep learning, several LV segmentation methods have been developed, achieving accurate results for myocardium extraction using MPS. For example, Ádám István Szűcs et al. employed a 3D U-Net model for self-supervised learning, achieving a Dice similarity coefficient (DSC) of 78.61% with limited labeled data [4]. Other models, such as the 3D V-Net developed by Wang et al., incorporate a composite loss function to optimize segmentation accuracy, reaching DSCs of 0.965 and 0.910 for epicardial and endocardial segmentation, respectively [5]. Similarly, the DP-ST-V-Net model proposed by Zhu et al. combines shape priors generated by dynamic programming with a spatial transformation network, significantly enhancing segmentation accuracy. This method achieved an average DSC of over 0.95 for endocardial, myocardial, and epicardial segmentation, while maintaining a Hausdorff distance of less than 8 mm, demonstrating robustness and accuracy across different severities of myocardial ischemia [6].

Despite these advancements, existing models rely heavily on centralized datasets, which are often limited to a single institution due to privacy concerns. In the medical field, image data from different hospitals cannot be shared directly due to strict privacy regulations, which limits the generalization of existing neural network models trained on centralized datasets. Federated learning, as a collaborative machine learning approach, addresses these challenges by enabling institutions to train a shared model without exchanging raw data, thus preserving patient privacy [7]. The Federated Averaging (FedAvg) algorithm proposed by McMahan et al. [8] aggregates model updates from participating sites, ensuring privacy protection while improving model performance and robustness. This decentralized approach allows institutions to collaboratively train models while maintaining the confidentiality of sensitive medical data.

However, federated learning introduces new challenges, particularly due to the inherent heterogeneity of medical imaging data across institutions [9]. In real-world applications, data distributions often vary significantly due to the differences in imaging equipment, protocols, and patient demographics. FedAvg assumes equal contribution of data from all sites, which can lead to suboptimal performance when data are non-iid (non-independent and identically distributed) [10]. Additionally, frequent communication between the central server and local devices can impose computational and bandwidth burdens, especially for resource-constrained institutions. To address these challenges, adaptive federated learning algorithms are needed to better manage data heterogeneity and ensure fairness across institutions, enabling models to achieve robust and equitable performance in diverse medical environments.

This paper introduces a novel deep learning-based method for LV segmentation, leveraging a **Fed**erated **D**omain **A**daptation with **T**ime**S**former (FedDA-TSformer) to extract both spatial and temporal features from 4D SPECT images using multi-centered MPS data, as shown in Figure 1. The federated learning and domain adaptation approaches are designed to utilize multi-center SPECT images while addressing data heterogeneity and preserving patient privacy. The input of FedDA-TSformer consists of a sequence of gated MPS, and the output is a probability map for the epicardial and endocardial masks. The proposed FedDA-TSformer was applied to MPS from three medical centers with a total of 150 patients and achieved a DSC of 0.907 for epicardium segmentation and 0.842 for endocardium segmentation, with average surface distances of 0.254 and 0.460 pixels, respectively. These results demonstrated superior performance compared to existing methods, highlighting the potential of the proposed method to achieve more accurate and robust LV segmentation in SPECT imaging while preserving the data privacy. This advancement ultimately supports improved functional assessment and accelerates the clinical diagnosis process.

The contributions of this paper are shown below:

1) We proposed a novel FedDA-TSformer model to perform feature representation learning using sequential gated MPS images for LV segmentation.

2) We leveraged federated learning to enable decentralized model training, ensuring patient data privacy while supporting multi-institutional collaboration.

3) We employed domain adaptation-related loss functions to address data heterogeneity and align feature representations into a homogeneous space while preserving data privacy.

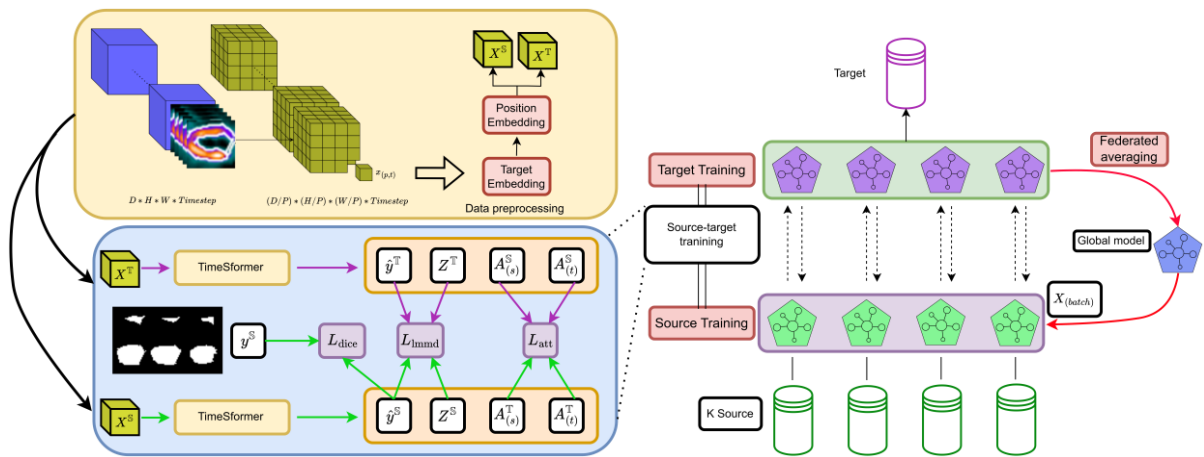

**Figure 1**. Overview of the proposed FedDA-TSformer for LV segmentation. First, the TimeSformer model is employed to perform feature extraction and extract the epicardium or endocardium using sequential MPS from a single hospital. Subsequently, federated learning is employed to train the model while maintaining data privacy; no data is exchanged between different hospitals, ensuring the data privacy. Meanwhile, during training, a combined objective function is employed to perform domain adaptation, addressing data heterogeneity and improving the model's generalizability.

## 2. Materials and Methods

### 2.1. Enrolled subjects and data acquisition

One-hundred and fifty fully de-identified MPS datasets from three hospitals were enrolled in this study. Within these 150 patients, 73 of them were from the First Affiliated Hospital of Nanjing Medical University (NJ), 28 patients were from Chang Bing Show Chwan Memorial Hospital, Taiwan (TW), and the rest 49 patients were from Xiangya Hospital, Central South University (XY). The detailed information of images acquisition from these three medical centers are shown below.

**NJ:** A total of seventy-three patients (47 males and 26 females; age range 38–83 years, mean age 62.18 ± 11.67 years) were retrospectively enrolled in this study. These patients underwent myocardial perfusion SPECT (MPS) imaging at Jiangsu Provincial Hospital, China, between March 2014 and August 2017. The study received approval from the hospital's Medical Ethics Committee. Each patient underwent 8 frames of ECG-gated rest and stress SPECT imaging following the injection of 20–30 mCi of Tc-99m sestamibi. In total, 1168 3D MPS image volumes (73 patients × 2 states [rest or stress] × 8 frames) were analyzed in this study. The SPECT images were reconstructed using an ordered subset expectation maximization (OSEM) algorithm with 3 iterations and 10 subsets, incorporating a Butterworth filter with a power of 10 and a cutoff frequency of 0.3 cycles/cm. Each SPECT image volume had a reconstructed voxel size of 6.4 mm × 6.4 mm × 6.4 mm. Based on medical conditions and MPS imaging findings at enrollment, the patients were categorized by a group of clinical cardiologists in consensus into three groups: 31 subjects without ischemia or with mild ischemia, 32 subjects with moderate ischemia, and 12 subjects with severe ischemia [11].

**TW**: The study protocol was approved by the ethics committee of Chang Bing Show Chwan Memorial Hospital, Taiwan and written informed consents were obtained from all patients. 28 patients were involved in the experiments [12]. One SPECT MPI image volumes from 8 phases were acquired in one cardiac cycle for each patient. Twenty 2D images were generated and stored in the long-axis form of heart by longitudinal sectioning with an interval of 9°, which helps locate the apex of LV.

**XY**: A total of forty-nine patients from Xiangya Hospital, Central South University, were enrolled in the study [13]. A routine resting gated myocardial perfusion imaging (MPI) scan was conducted 60 minutes after intravenous administration of 925–1110 MBq technetium (Tc)-99m sestamibi (supplied by the Chinese Atomic Energy Institute, Changsha, China). The scan was performed using a dual-camera SPECT system (Discovery 630, GE, USA) with a 20% energy window centered at 140 keV. The gated imaging was triggered by the ECG R-wave, and each cardiac cycle was divided into 8 frames. A total of 64 planar projections were obtained (32 steps with 25 seconds per step) using the step-and-shoot mode, in accordance with the guidelines recommended by the American Society of Nuclear Cardiology (ASNC) [14]. Image reconstruction and reorientation were completed with Emory Reconstruction Toolbox (ERToolbox; Atlanta, GA). MPS data was reconstructed by OSEM with three iterations and 10 subsets and then filtered by a Butterworth lowpass filter with a cut-off frequency of 0.4 cycles/cm and an order of 10.

**Data preparation**. Long-axis slices are more effective than short-axis slices for identifying apical and basal positions. To process each 3D MPS image volume, it was initially divided longitudinally to produce 32 2D long-axis slices 32, each with a resolution of 32 × 32 [15]. Since regions far from the LV in long-axis slices provide minimal information for LV segmentation, we cropped the central 32 pixels from each slice, resulting in 32 cropped long-axis slices and 32 short-axis slices, both with a resolution of 32 × 32. Consequently, each scanned gate (cardiac frame) produced a 3D volume sized 32 × 32 × 32. The data generation workflow is illustrated in Figure 2. An experienced nuclear cardiologist manually delineated the epicardial and endocardial contours for each gate of the MPS, and the results were validated by a senior nuclear cardiologist (Z.X and G.H). These contours were then converted into binary masks to serve as manual reference annotations.

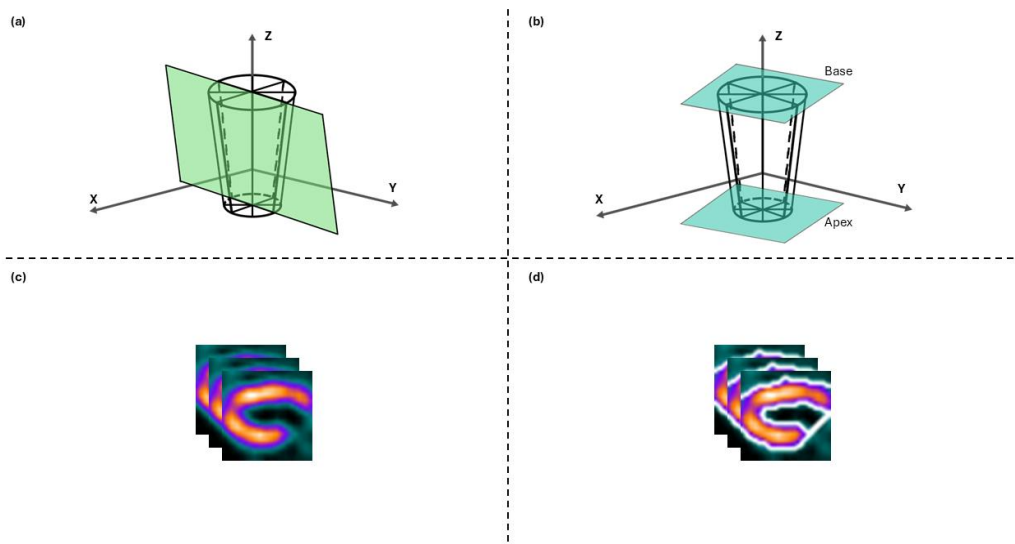

**Figure 2** Pre-processing workflow for LV MPS. (a) The 3D MPS volume is segmented into 32 long-axis slices by slicing it along the LV's longitudinal direction (z-axis), as illustrated by the green plane. (b) The myocardial regions, extending from the LV base to the apex, are identified and segmented using the vertical blue planes. (c) An example of a long-axis slice from the MPS image is provided for reference. (d) Endocardial and epicardial contours are manually annotated on the long-axis slice, as shown in white curves.

## 2.2 TimeSformer for LV segmentation

By stacking MPS slices, we obtain a 3D MPS volume with the size of $H \times W \times D$, where $H$, $W$, and $D$ represent the height, width, and depth of the volume. The LV segmentation mask is a 3D volume where the value 1 indicates the epicardium or endocardium, and 0 represents the background. We reformulate the LV segmentation problem as a time-series prediction task using the spatial-temporal information of the 4D MPS. The input of our model is denoted as $X \in \mathbb{R}^{H \times W \times D \times 1 \times T}$, consists of a series of gated MPS 3D volumes, where 1 indicates that the MPS image is one-channel image and $T$ is the number of gates processed for LV segmentation. Mathematically, the LV segmentation problem is denoted as the epicardial or endocardial prediction task while considering the spatial-temporal MPS for the $T$ consecutive frames.

The temporal dimension is especially significant because the gated MPS frames represent dynamic changes in the myocardium over time. Traditional models may struggle to handle these complex spatial-temporal interactions, particularly in scenarios where both fine-grained spatial details and long-range temporal dependencies need to be analyzed simultaneously. Thus, we employ the TimeSformer model [16] for its ability to effectively capture both spatial and temporal dependencies, which are crucial for accurate LV segmentation in 4D MPS data. The TimeSformer model's attention mechanism is well-suited for addressing these challenges, enabling the model to focus on critical features across both spatial and temporal dimensions, thereby improving segmentation accuracy and consistency across frames.

However, the TimeSformer model is designed for 2D image prediction, while our input is 3D, and we require a 3D output. To address this, we modified the TimeSformer model to take 3D input and produce 3D predictions. Following the TimeSformer, we decompose each gate into patches of size $P \times P \times P$, resulting in $N$ patches for the 3D volume, where $N = HWD/P^3$. This decomposition reduces computational complexity while preserving essential local information, enabling efficient and effective processing of 3D data [17]. Then, these patches are flattened into vectors $x_{(p,t)} \in \mathbb{R}^{1 \times P \times P \times P}$, with $p \in \{1, \dots, N\}$ representing spatial locations, and $t \in \{1, \dots, T\}$ indexing over the gates. The overview of the proposed TimeSformer is shown in Figure 3.

Each patch $x_{(p,t)}$ is linearly transformed into an embedding vector $z_{(p,t)}^{(0)} \in \mathbb{R}^D$ using a learnable matrix $E \in \mathbb{R}^{D \times 1P^3}$, as denoted in Eq. 1.

$$z_{(p,t)}^{(0)} = E x_{(p,t)} + e_{(p,t)}^{pos} \tag{1}$$

where $e_{(p,t)}^{pos} \in \mathbb{R}^D$ is a learnable positional embedding added to encode the spatiotemporal information of each patch. The final sequence of embedding vectors $z_{(p,t)}^{(0)}$ is computed for spatial locations $p \in \{1, \dots, N\}$ and for time steps $t \in \{1, \dots, T\}$.

To capture the contextual information from the sequential MPS, we introduce a special learnable vector $z_{(0,0)}^{(0)} \in \mathbb{R}^D$, which serves as the embedding for the classification token, CLS token acts as a global descriptor that aggregates contextual information from the entire sequence. In detail, the employed TimeSformer consists of $L$ encoding blocks. In each block $\ell \in \{1, \dots, L\}$, a query, key, and value vector are computed for each patch based on the representation $z_{(p,t)}^{(\ell-1)}$, which is encoded by the preceding block, denoted as $\ell - 1$ which is the previous block result. In detail, the query, key and values are denoted in Eqs. 2 to 4.

$$q_{(p,t)}^{(\ell,a)} = W_Q^{(\ell,a)} \text{LN}\left(z_{(p,t)}^{(\ell-1)}\right) \in \mathbb{R}^{D_h} \tag{2}$$

$$k_{(p,t)}^{(\ell,a)} = W_K^{(\ell,a)} \text{LN}\left(z_{(p,t)}^{(\ell-1)}\right) \in \mathbb{R}^{D_h} \tag{3}$$

$$v_{(p,t)}^{(\ell,a)} = W_V^{(\ell,a)} \text{LN}\left(z_{(p,t)}^{(\ell-1)}\right) \in \mathbb{R}^{D_h} \quad (4)$$

where $LN(\cdot)$ represents the LayerNorm [18], aiming at stabilizing and accelerating the training process by normalizing the inputs across features within each layer. This helps to mitigate issues related to internal covariate shift, ensures faster convergence, and improves generalization by maintaining consistent activation distributions throughout the network [18]. In these equations, $a \in \{1, \cdots, A\}$ indicates the index of the multiple attention heads, where $A$ is the total number of attention heads. The latent dimensionality for each attention head is set to $D_h = D/A$.

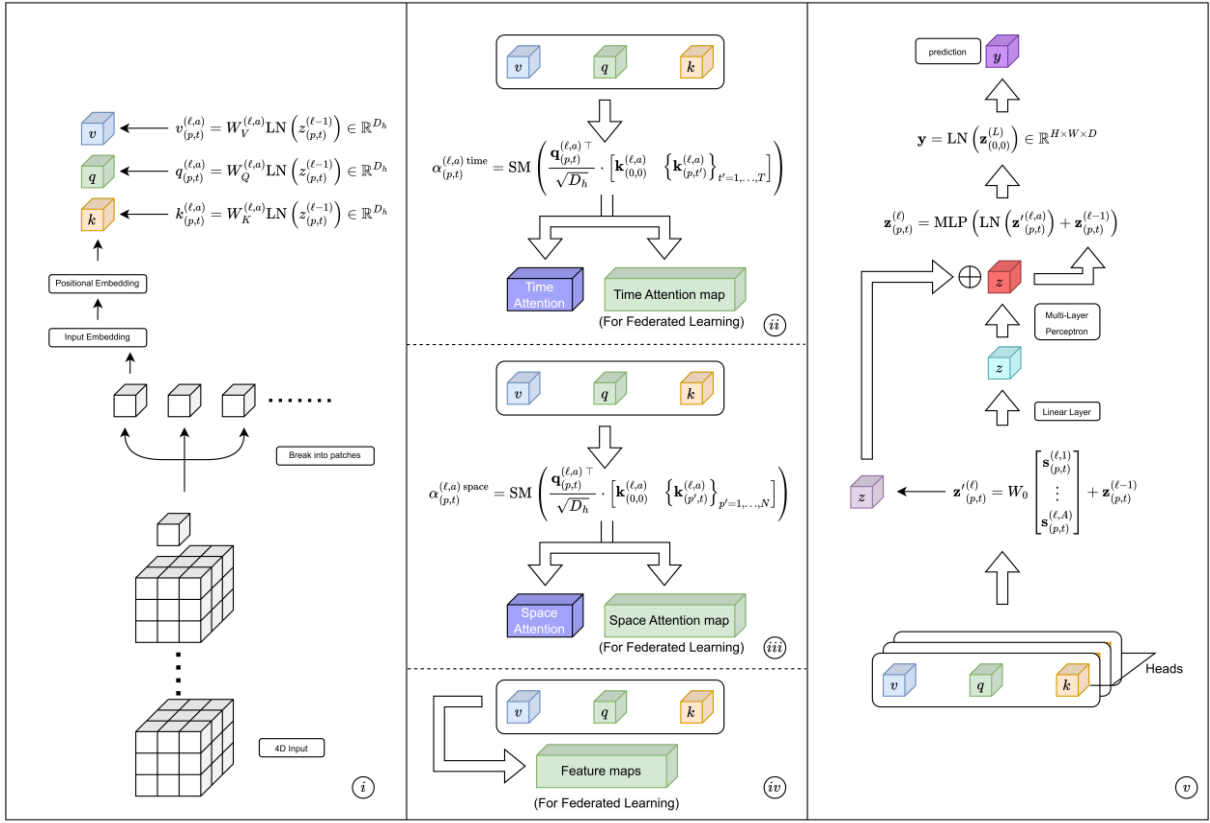

**Figure 3.** The TimeSformer workflow follows a numbered sequence in the lower right corner, which guides the process. It begins with a 4D input that is divided into small patches. These patches are then processed through input embedding and position embedding layers, followed by Q, K, and V encoding blocks. The second, third, and fourth stages focus on temporal self-attention computation, spatial self-attention computation, and the extraction of feature maps, respectively. During these stages, the Time Attention map, Space Attention map, and Feature maps are extracted but not used in the computation, as they are prepared for later integration into the FedDA-TSformer framework. Finally, in the fifth stage, the results of the multi-head computations are integrated to generate the final prediction.

LV segmentation from 4D MPS involves analyzing dynamic cardiac activity across a sequence of frames while simultaneously identifying intricate spatial features within each frame. This dual dependency—on spatial detail and temporal evolution—necessitates a mechanism capable of capturing correlations both within and across frames. Instead of jointly attending across both time and space dimensions, as in Joint Space-Time Attention [16], we employ Divided Space-Time Attention [16] to perform spatial and temporal attention for improved feature representation learning and reduced computational complexity. Divided Space-Time Attention first computes attention along the temporal axis (capturing relationships across frames) and then computes attention along the spatial axis (capturing relationships within each frame), as shown in Figure 4. These two computations are subsequently combined, enabling the integration of both temporal and spatial correlations in a computationally efficient manner. Specifically, the computational complexity of Divided Space-Time Attention is: $O((N+T)^2)$ in contrast to Joint Space-Time Attention $O(T^2 \cdot N + N^2 + T)$. As a result, we reduced the complexity while preserving the ability to model long-range dependencies in both temporal and spatial domains in LV segmentation task using 4D MPS.

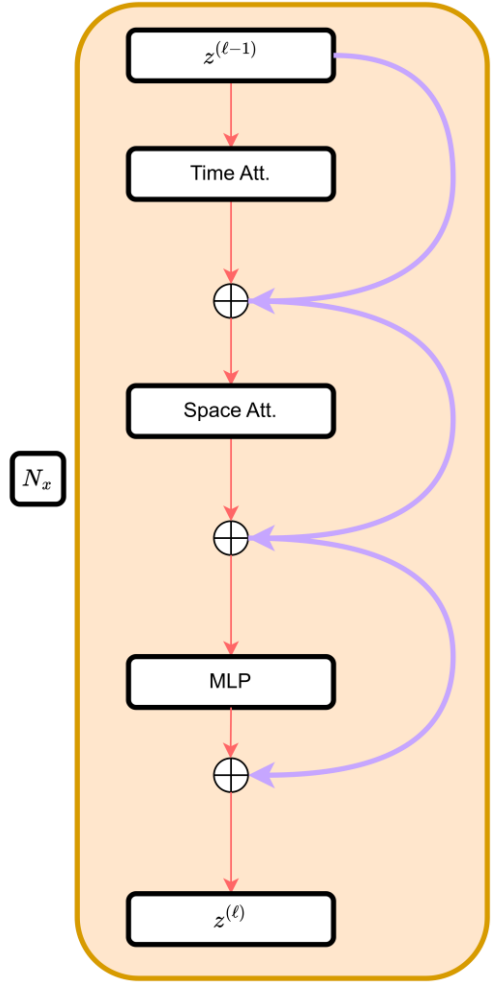

**Figure 4**. Architecture of the Divided Space-Time Attention block implemented in our work. Self-attention is applied across all blocks, with residual connections represented by the purple line. $N_x$ indicates the block depth, referring to how many times the structure has been stacked.

In detail, suppose the self-attention weights $\alpha_{(p,t)}^{(\ell,a)\,time+space} \in \mathbb{R}^{N+G+2}$ for a query patch $(p,t)$ are denoted in Eqs. 5 and 6.

$$\alpha_{(p,t)}^{(\ell,a)\,time} = \text{SM}\left( \frac{q_{(p,t)}^{(\ell,a)\,\top}}{\sqrt{D_h}} \cdot \left[ k_{(0,0)}^{(\ell,a)} \left\{ k_{(p,t')}^{(\ell,a)} \right\}_{t'=1,\ldots,T} \right] \right) \quad (5)$$

$$\alpha_{(p,t)}^{(\ell,a)\,space} = \text{SM}\left( \frac{q_{(p,t)}^{(\ell,a)\,\top}}{\sqrt{D_h}} \cdot \left[ k_{(0,0)}^{(\ell,a)} \left\{ k_{(p',t)}^{(\ell,a)} \right\}_{p'=1,\ldots,N} \right] \right) \quad (6)$$

The encoding $z_{(p,t)}^{(\ell)}$ at block $\ell$ is obtained by first calculating the weighted sum of the value vectors using the self-attention coefficients from each attention head. However, since we are using Divided Space-Time Attention, the computation is split and performed as shown in Eq. 7.

$$s_{(p,t)}^{(\ell,a)} = \alpha_{(p,t),(0,0)}^{(\ell,a)} v_{(0,0)}^{(\ell,a)} + \sum_{t'=1}^{G} \alpha_{(p,t),(p,t')}^{(\ell,a)} v_{(p,t')}^{(\ell,a)} + \sum_{p'=1}^{N} \alpha_{(p,t),(p',t)}^{(\ell,a)} v_{(p',t)}^{(\ell,a)} \tag{7}$$

Next, the concatenated vectors from all attention heads are projected and fed into a multilayer perceptron (MLP), with residual connections applied after each operation, as shown in Eqs. 8 and 9.

$$z'^{(\ell)}_{(p,t)} = W_O \begin{bmatrix} s_{(p,t)}^{(\ell,1)} \\ \vdots \\ s_{(p,t)}^{(\ell,\mathcal{A})} \end{bmatrix} + z_{(p,t)}^{(\ell-1)} \tag{8}$$

$$z_{(p,t)}^{(\ell)} = \text{MLP}\left(\text{LN}\left(z'^{(\ell)}_{(p,t)}\right)\right) + z_{(p,t)}^{(\ell-1)} \tag{9}$$

The final clip embedding is extracted from the last block, specifically from the classification token.

$$y = MLP\left(LN\left(z_{(0,0)}^{(L)}\right)\right) \in \mathbb{R}^{H \times W \times D} \tag{10}$$

where $y \in \mathbb{R}^{H \times W \times D}$ indicates the final predicted endocardium or epicardium masks. A 1-hidden-layer MLP is added on top of this representation to predict the final image segmentation. In this case, the output dimensions are $32 \times 32 \times 32$.

### 2.3 FedDA-TSformer: Privacy-Preserving Approach for LV Segmentation

Medical imaging data, such as 4D MPS, is highly sensitive, and sharing raw patient data between institutions raises significant privacy and compliance concerns [19]. With three distinct centers involved in this study, it is critical to develop a strategy that enables collaborative training of a robust segmentation model while ensuring that patient data remains confidential and adheres to data protection regulations, such as HIPAA and GDPR [20]. Traditional centralized training approaches, which require pooling data from all centers, are not feasible due to these privacy constraints and the potential legal and ethical implications of sharing patient information [21]. Additionally, each center may have unique data distributions due to variations in equipment, imaging protocols, and patient demographics, further complicating centralized approaches. In the context of this study, the source domain refers to the dataset obtained from one of the three participating centers, while the target domain indicates the datasets from the other centers used for testing and evaluating the model's generalizability.

Federated learning ensures client privacy by keeping local data and identifiable information on the client side, without transmitting them over the network [22]. Instead, only gradients and model weights are shared during the training process. Compared to training methods, which require data from both the source and target sites to be exchanged and used, federated learning protects patient information by avoiding any data exchange between sites. A baseline method in federated learning is FedAvg [8], which operates without accessing raw data, relying solely on gradients and model weights for model updates. The aggregation formula is expressed as in Eq. 11.

$$\theta = \sum_{i=0}^{N} \left(\frac{S_i}{S} \theta_i\right) \tag{11}$$

where $N$ represents the number of sites (hospitals in LV segmentation task), $S$ is the total number of samples involved in the training, $S_i$ is the sample size of the $i$-th client, and $\theta_i$ represents the model weights for $i$-th client. The server distributes the updated global model to all clients, and the process repeats until the model converges. Throughout this process, the original data remains stored locally, ensuring the protection of data privacy.

However, the distributions of MPS images from these three hospitals are different, and the performance of models may drop rapidly in the target domain [23]. However, as mentioned in Section 2.1, these datasets vary in characteristics such as imaging protocols, scanner types, and patient demographics, collectively representing diverse input distributions. The target domain, on the other hand, represents another dataset from one of these three centers that differs from the source domain in terms of acquisition settings or population characteristics. The goal is to ensure that the trained model can generalize effectively to this new domain despite the inherent domain shift, while keeping the data privacy.

To enhance the generalization ability of the model in the target domain, the data of the target domain is added for domain adaptation learning. We implement our federated learning approach using along with the TimeSformer. Notably, two distinct self-attention maps are produced as part of this process, as shown in Figure 5.

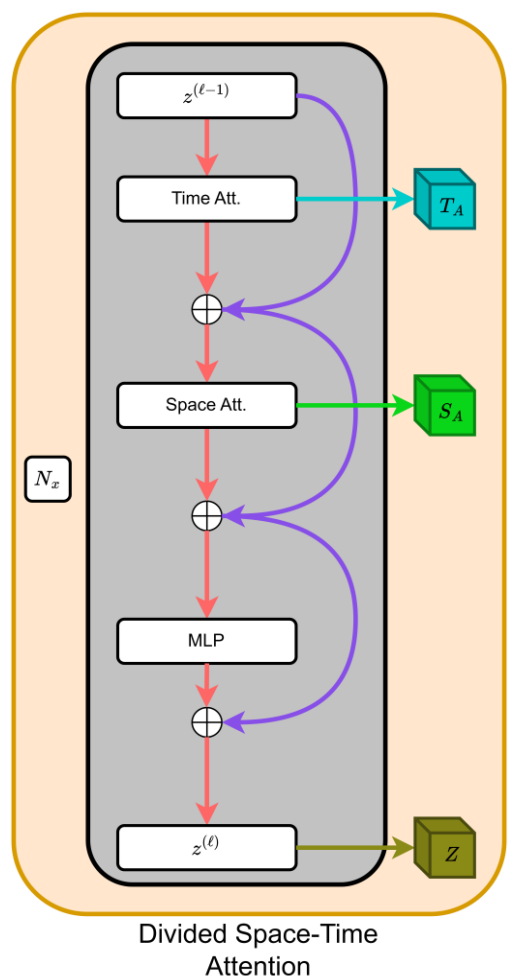

**Figure 5**. Overview of the modified Divided Space-Time Attention structure to enable the collection of attention maps and feature maps, making it adaptable for use with FedDA-TSformer.

In Figure 5, we introduce $T_A$ and $S_A$, where

- $T_A$ represents the **Time Attention Map**, responsible for capturing temporal dependencies, as shown in Eq. 12.

$$T_A = \text{SM}\left(\frac{{q^{(\ell,a)}_{(p,t)}}^\top \cdot k^{(\ell,a)}_{(p,t')}}{\sqrt{D_h}}\right) \qquad (12)$$

- $S_A$ represents the **Spatial Attention Map**, which captures spatial relationships, as shown in Eq. 13.

$$S_A = \text{SM}\left(\frac{{q^{(\ell,a)}_{(p,t)}}^\top \cdot k^{(\ell,a)}_{(p',t)}}{\sqrt{D_h}}\right) \qquad (13)$$

Additionally, $Z$ represents the feature encoded by TimeSformer.

### 2.4 Optimization and loss function

When working with data from both source and target domains, significant domain shifts—caused by differences in imaging protocols, patient populations, or equipment—can lead to inconsistencies in the model's performance. Addressing these discrepancies requires a targeted approach that aligns both the spatial-temporal attention patterns and the class-wise feature distributions across the two domains. In this paper, we employ Self Attention (SA) consistency loss and local maximum mean discrepancy (LMMD) loss to perform federated learning while performing the domain adaptation.

**SA Consistency Loss**: The need for SA consistency loss arises from the importance of ensuring that the self-attention mechanisms of the model capture similar spatial and temporal relationships in both domains. Differences in attention distributions between the source and target domains can lead to misalignment in how the model interprets features, particularly for critical structures such as the epicardium and endocardium. By enforcing consistency between the self-attention maps, we ensure that the model learns robust spatial-temporal representations that are invariant to domain-specific variations.

We apply SA Consistency Loss to ensure that the self-attention maps generated by the transformer for both the source and target domains remain consistent. Specifically, we extract the SA matrices from the last layer of the encoder for each sample, because it can capture more abstract and semantically meaningful relationships compared to earlier layers. And from both spatial attention and temporal attention in Figure 5. For the time attention map, we have $T_A \in \mathbb{R}^{(V/P)^3 HB \times Token \times (H+Token)}$ and for the spatial attention map, we have $S_A \in \mathbb{R}^{HB \times (V/P)^3 \times (V/P)^3 H}$. Here, $V$ denotes the volume of the input data, $P$ represents the patch size, $H$ refers to the number of attention heads, $B$ is the batch size, and Token refers to the sequence length. To ensure that the attention distributions between the source and target domains are aligned, we transform and reshape the attention matrices to make them comparable. Once aligned, we minimize their difference using the consistency loss.

Based on the FedDAvT framework, it originally uses a single attention map with one corresponding loss. However, since we have two distinct attention maps, we modify the SA consistency loss into the following two components:

For the time attention map, the objective function is defined in Eq. 14.

$$\mathcal{L}^T_{att} = \frac{1}{(V/P)^3 HB \times Token \times (H+Token)} \left\|T_A^{\mathbb{S}} - T_A^{\mathbb{T}}\right\| \qquad (14)$$

For the space attention map, the objective function is defined in Eq. 15.

$$\mathcal{L}^S_{att} = \frac{1}{HB \times (V/P)^3 \times (V/P)^3} \left\|S_A^{\mathbb{S}} - S_A^{\mathbb{T}}\right\| \qquad (15)$$

where $\mathbb{T}$ represents that it is from the target domain, and $\mathbb{S}$ represents that it is from the source domain, ensuring alignment between the attention distributions of the source and target domains. At the end, we combine the two components—time attention loss and space attention Loss—to compute the final SA consistency Loss. The final SA consistency loss is formulated as in Eq. 16.

$$\mathcal{L}_{att} = \mathcal{L}_{att}^T + \mathcal{L}_{att}^S \tag{16}$$

This combined loss encourages both temporal and spatial attention alignment across the domains, improving the model's ability to generalize effectively in domain adaptation tasks.

**LMMD Loss**: The objective of the LMMD loss is to facilitate domain adaptation by minimizing the distribution discrepancy between the source domain and the target domain, with a specific focus on category-level alignment. Unlike traditional MMD, which focuses on global distribution matching, LMMD addresses class-level misalignments that are critical for accurate segmentation. By ensuring that the features of each class (e.g., myocardium versus background) in the source domain are closely aligned with their counterparts in the target domain, LMMD enables more precise domain adaptation. This class-specific alignment is particularly effective in scenarios where target labels are unavailable, enhancing the model's ability to generalize and produce accurate segmentations in the target domain. The LMMD loss is formulated as in Eq. 17.

$$\mathcal{L}_{lmmd}(Z^{\mathbb{S}}, Z^{\mathbb{T}}) = \frac{1}{c}\sum_{c=0}^{c}\left\|\sum_{i=1}^{N^{\mathbb{S}}}\omega_i^{\mathbb{S}c}\phi(Z_i^{\mathbb{S}}) - \sum_{j=1}^{N^{\mathbb{T}}}\omega_i^{\mathbb{T}c}\phi(Z_j^{\mathbb{T}})\right\|_{\mathcal{H}}^2 =$$
$$\frac{1}{c}\sum_{c=1}^{c}\left[\sum_{i=1}^{N^{\mathbb{S}}}\sum_{j=1}^{N^{\mathbb{S}}}\omega_i^{\mathbb{S}c}\omega_j^{\mathbb{S}c}k(Z_i^{\mathbb{S}}, Z_j^{\mathbb{S}}) + \sum_{i=1}^{N^{\mathbb{T}}}\sum_{j=1}^{N^{\mathbb{T}}}\omega_i^{\mathbb{T}c}\omega_j^{\mathbb{T}c}k(Z_i^{\mathbb{T}}, Z_j^{\mathbb{T}}) - 2\sum_{i=1}^{N^{\mathbb{S}}}\sum_{j=1}^{N^{\mathbb{T}}}\omega_i^{\mathbb{S}c}\omega_j^{\mathbb{T}c}k(Z_i^{\mathbb{S}}, Z_j^{\mathbb{T}})\right] \tag{17}$$

where $\mathcal{H}$ denotes the Reproducing Kernel Hilbert Space (RKHS), and $\phi(\cdot)$ is the mapping function that transforms the original samples into the RKHS. The weight $\omega_i^c$ is computed as follows:

$$\omega_i^c = \frac{y_{ic}}{\sum_{y_i \in y} y_{ic}} \tag{18}$$

where $y_{ic}$ is the label that the sample belongs to each category. However, only the source domain $\mathbb{S}$ can provide labels, while the target domain $\mathbb{T}$ has no labels. Therefore, the pseudo labels $\hat{y}_{ic}$ generated by the classification layer need to be used. Finally, the objective function of the whole learning is denoted in Eq. 19.

$$\mathcal{L} = \mathcal{L}_{Dice} + \alpha * \mathcal{L}_{att} + \beta * \mathcal{L}_{lmmd} \tag{19}$$

Here, $\alpha$ and $\beta$ serve as the balancing factors in the domain adaptation loss, allowing us to fine-tune the contributions of the attention loss and the LMMD loss.

By combining these loss functions, the framework addresses both spatial-temporal attention alignment and feature distribution alignment, resulting in a model that is both robust and capable of effective generalization across domains.

**Optimization**: We employ the Federated Averaging method (FedAvg) to aggregate the model weights and gradients while considering the domain adaptation using Eq. 19. The process proceeds as follows:

1) The server initializes the global model and sends it to the clients.
2) The clients train the model locally using their own data using Eq. 19.
3) After training, the clients send their updated model weights to the server.
4) The server aggregates the local model updates using FedAvg, computes the new global model, and updates the gradients accordingly.

By implementing FedDAvT, our model is capable of performing federated learning in a multi-source domain adaptation setting. This approach ensures the privacy of patient data.

### 2.5 Evaluation metrics

The proposed LV myocardium segmentation method consists of a training stage and a testing stage. A retrospective dataset with 150 subjects were enrolled and each subject contains eight gates of the gated MPS. As a result, 1200 volumes with the size of 32 × 32 × 32 were generated and each volume represents a dedicated training sample. We employed five-fold cross-validation to train and test models. Each fold includes 120 subjects, sampled proportionally based on the distribution of subjects from the three hospitals, as the training set. It is important to note that the data splitting process is based on the subject level; in other words, the gated MPS data for the same subject was not divided into separate training and testing sets.

To evaluate the model performance, we first checked the generated contours of the epicardium and endocardium visually. To quantitatively reflect the voxel difference, we adopted DSC, sensitivity (SN), and specificity (SP) to evaluate the model performance. The SN measures the proportion of the true positive (TP) voxels compared to the number of TP and false negative (FN) voxels, while the SP measures the proportion of true negative (TN) voxels compared to the number of TN and false positive (FP) voxels. DSC, SN, and SP of 1 indicate a perfect match between the model prediction and the manual annotation; on the contrary,0 represents a total mismatch.

Additionally, we used Hausdorff Distance (HD) and Average Surface Distance (ASD) to evaluate the geometric similarity between the predicted and ground truth contours. HD calculates the maximum surface distance between the predicted and reference boundaries, reflecting the worst-case error, while ASD computes the average surface-to-surface distance, offering a more comprehensive measure of contour alignment.

## 3. Experimental results

### 3.1 Baseline Models

To evaluate the effectiveness of the proposed model, we use V-Net, a foundational 3D segmentation model, as the baseline. V-Net extracts hierarchical spatial features through convolutional layers, max-pooling, and up-sampling. Additionally, we incorporate two federated learning methods—FedAvg and FedDAvT—to enable collaborative model training while preserving data privacy across sites and to compare the performance of these approaches.

### 3.2 Results of myocardium segmentation

We evaluated our model using different MPS gates as sequential inputs. **Table 1** presents a quantitative comparison of the baseline model's performance with the proposed FedDA-TSformer for LV epicardial and endocardial segmentation.

**Table 1**. Quantitative evaluation for the V-Net, V-Net with FedAvg, TimeSformer, TimeSformer with FedAvg and the proposed **FedDA-TSformer** for LV epicardium and endocardium segmentation. The evaluation metrics include Dice Similarity Coefficient (DSC), Hausdorff Distance (HD) in pixels, Average Surface Distance (ASD) in pixels, Sensitivity (SN), and Specificity (SP). The number of gates refers to the length of the MPS sequence used to achieve the best performance. The bold text highlights our proposed model. Arrows (↑) denote that higher values are 1preferable, while arrows (↓) indicate that lower.

| Class | Model | # Gates | DSC↑ | HD↓ | ASD↓ | SN↑ | SP↑ |
|---|---|---|---|---|---|---|---|
| Epicardium | V-Net | 2 | 0.923±0.02 | 7.090±2.187 | 0.195±0.064 | 0.919±0.047 | 0.933±0.038 |
| | V-Net-FedAvg | 2 | 0.922±0.02 | 8.161±2.147 | 0.201±0.063 | 0.925±0.051 | 0.922±0.043 |
| | TimeSformer | 2 | 0.915±0.02 | 8.491±2.221 | 0.224±0.073 | 0.925±0.045 | 0.903±0.060 |
| | TimeSformer-Fed | 2 | 0.916±0.023 | 8.477±2.366 | 0.222±0.088 | 0.922±0.055 | 0.910±0.063 |
| | FedDA-TSformer | 2 | 0.907±0.030 | 8.521±2.329 | 0.254±0.135 | 0.918±0.064 | 0.9±0.064 |
| Endocardium | V-Net | 4 | 0.803±0.062 | 8.999±1.916 | 0.532±0.190 | 0.867±0.062 | 0.959±0.018 |
| | V-Net-FedAvg | 4 | 0.803±0.62 | 9.232±2.142 | 0.538±0.199 | 0.866±0.068 | 0.961±0.020 |
| | TimeSformer | 2 | 0.834±0.02 | 8.560±2.098 | 0.444±0.179 | 0.853±0.045 | 0.970±0.061 |
| | TimeSformer-Fed | 2 | 0.831±0.056 | 8.511±2.214 | 0.460±0.223 | 0.854±0.064 | 0.970±0.011 |
| | FedDA-TSformer | 2 | 0.842±0.046 | 8.324±2.003 | 0.408±0.151 | 0.864±0.046 | 0.970±0.012 |

As shown in **Table 1**, our proposed FedDA-TSformer achieved the highest DSC and SP, along with the lowest ASD and HD, on the combined dataset from three sources, using 80% of the data for training and 20% for testing. The epicardial and endocardial contours generated by different models for representative slices are illustrated in **Figure 6**.

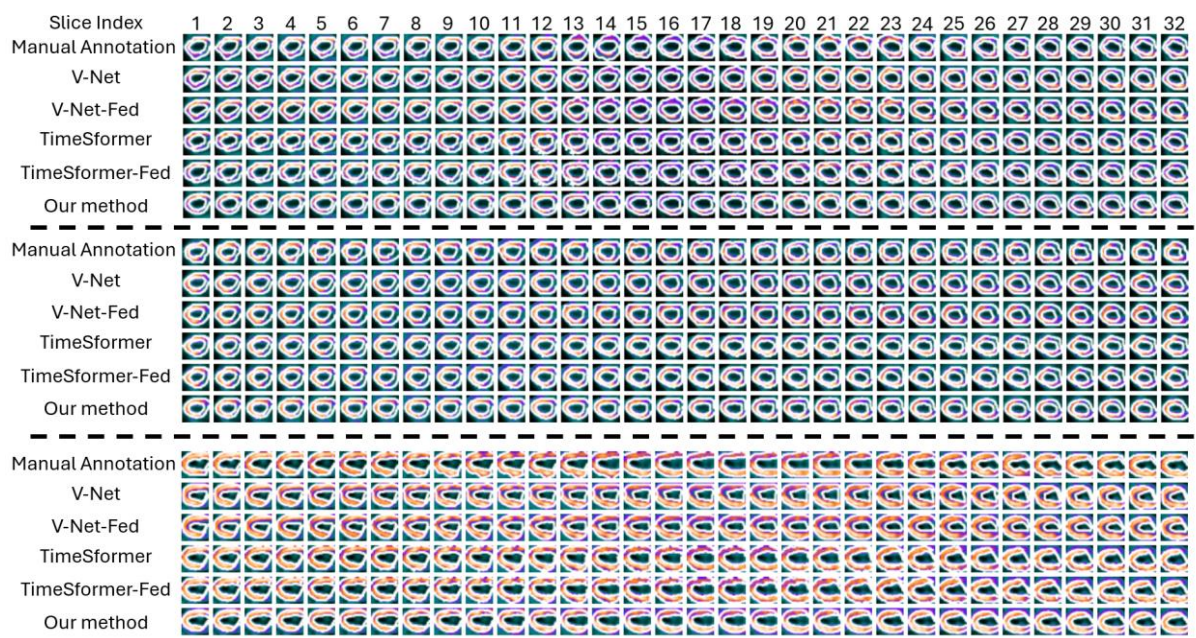

**Figure 6.** Comparison of the segmentation results for different models in the axial view. The top row corresponds to patient #41, gate #1; the middle row to patient #24, gate #2; and the bottom row to patient #53, gate #1. Each grouped image presents the epicardial and endocardial contours generated by manual annotation, V-Net, V-Net-Fed, TimeSformer, TimeSformer-Fed, and FedDA-TSformer for comparison.

As illustrated in Figure 6, a comparative analysis between the various models and manual segmentation reveals that our model exhibits a stronger temporal correlation compared to its counterparts. While it does not perfectly align with the manual segmentation, it consistently preserves the dynamic regions of segmentation that evolve over time.

We further validated the performance of the proposed FedDA-TSformer and the baseline TimeSformer model, both with and without FedAvg, using MPS sequences with varying numbers of gates. The results are presented in Figure 7.

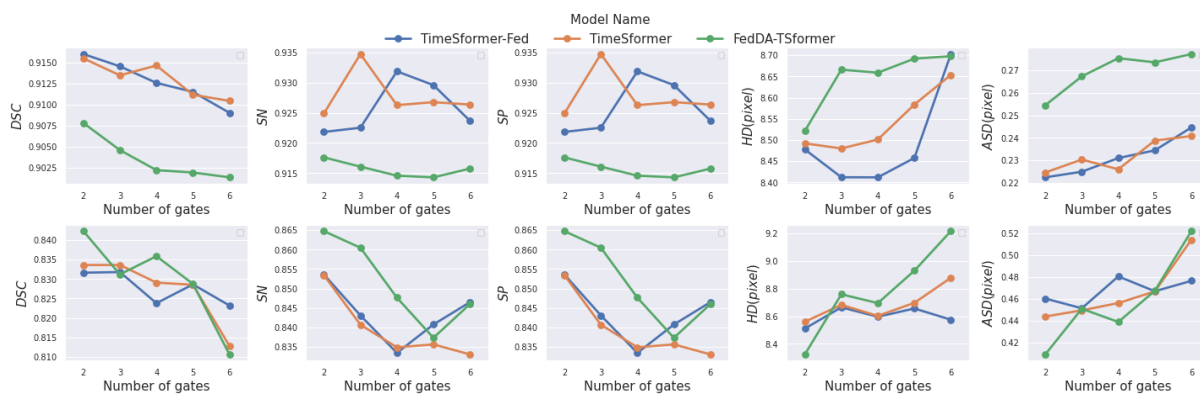

**Figure 7**. LV segmentation using different numbers of gates on different models (TimeSformer, TimeSformer+FedAVG, FedDA-TSformer). Draw a figure that the horizontal axis indicates the number of enrolled gates, the vertical axis indicates different metrics, including DSC, SN, SP, HD and ASD. Like the figure above.

According to Figure 7, FedDA-TSformer achieves the best performance when using the MPS sequence with two gates for both endocardium and epicardium segmentation. Specifically, the FedDA-TSformer performs the best on endocardium when using the two-gate MPS sequence, outperforming both the models with and without FedAvg. However, for epicardium segmentation, its performance is slightly lower compared to the other two models.

Eight gated MPS images were evaluated to assess the segmentation performance of various models: V-Net, V-Net-Fed, TimeSformer, TimeSformer with FedAvg, and our proposed FedDA-TSformer. The results are presented in **Figure 8**.

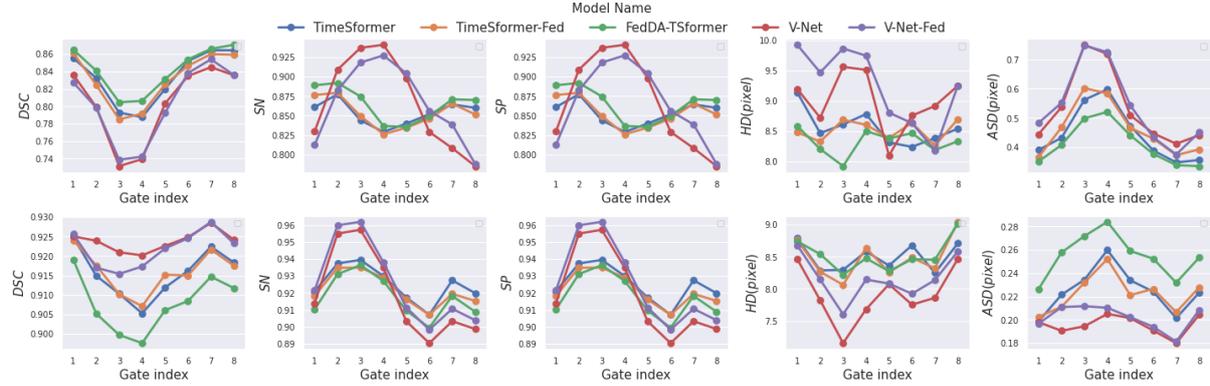

**Figure 8.** Quantitative evaluation of RV epicardium and endocardium segmentation for each MPS gate is provided. The top section presents results for epicardium segmentation, while the bottom section shows endocardium segmentation. Metrics include DSC (Dice Similarity Coefficient), HD (Hausdorff Distance in pixels), ASD (Average Surface Distance in pixels), SN (Sensitivity), and SP (Specificity).

Figure 8 demonstrates that our model consistently outperforms others in DSC metrics for endocardium segmentation across all gates, while maintaining performance comparable to the average in other aspects.

### 3.3. Cross-site performance comparison

To evaluate the multi-site performance of FedDA-TSformer across NJ, TW, and XY datasets, a model was trained using data from two hospitals with two gates and tested on the remaining site. The results are summarized in **Table 2**.

We have three hospitals, NJ, TW and XY. Build a model with 2 gates on data from 2 hospital, and evaluate the performance on the rest one.

**Table 2**. presents the performance of the TimeSformer baseline model and the FedDA-TSformer model using cross-site datasets from NJ (Nanjing Medical University), TW (Chang Bing Show Chwan Memorial Hospital, Taiwan), and XY (Xiangya Hospital, Central South University). In each experiment, two datasets were combined as the training set, while the remaining dataset was used as the testing set. The evaluation includes both endocardium and epicardium segmentation.

| Model | Contour Type | Training set | Testing set | DSC↑ | HD(pixel)↓ | ASD(pixel)↓ | SN↑ | SP↑ |
|---|---|---|---|---|---|---|---|---|
| TimeSformer | Endo | NJ+TW | XY | 0.7425 | 9.4018 | 0.8332 | 0.6798 | 0.9714 |
|  | Endo | NJ+XY | TW | 0.7378 | 11.4720 | 0.9459 | 0.8638 | 0.9410 |
|  | Endo | TW+XY | NJ | 0.7131 | 13.8711 | 1.1276 | 0.6846 | 0.9634 |
| TimeSformer+ FedAVG | Endo | NJ+TW | XY | 0.7315 | 10.6074 | 0.9723 | 0.7087 | 0.9549 |
|  | Endo | NJ+XY | TW | 0.7319 | 12.7287 | 1.0124 | 0.8550 | 0.9396 |
|  | Endo | TW+XY | NJ | 0.7140 | 13.9628 | 1.1613 | 0.6744 | 0.9666 |
| FedDA-TSformer | Endo | NJ+TW | XY | 0.7721 | 10.7829 | 0.7906 | 0.8017 | 0.9439 |
|  | Endo | NJ+XY | TW | 0.7515 | 10.0237 | 0.7688 | 0.8946 | 0.9443 |
|  | Endo | TW+XY | NJ | 0.7329 | 13.4874 | 0.9988 | 0.6748 | 0.9751 |
| TimeSformer | Epi | NJ+TW | XY | 0.9057 | 10.8322 | 0.2812 | 0.8837 | 0.9125 |
|  | Epi | NJ+XY | TW | 0.8925 | 10.7913 | 0.3242 | 0.8336 | 0.9310 |
|  | Epi | TW+XY | NJ | 0.8664 | 10.9141 | 0.4301 | 0.9473 | 0.7771 |
| TimeSformer+ FedAVG | Epi | NJ+TW | XY | 0.9032 | 10.7737 | 0.2930 | 0.8966 | 0.8812 |
|  | Epi | NJ+XY | TW | 0.8581 | 12.1008 | 0.4710 | 0.7637 | 0.9678 |
|  | Epi | TW+XY | NJ | 0.8666 | 11.0312 | 0.4325 | 0.9464 | 0.7771 |
| FedDA-TSformer | Epi | NJ+TW | XY | 0.8991 | 11.1468 | 0.3050 | 0.8660 | 0.9231 |
|  | Epi | NJ+XY | TW | 0.8531 | 11.4952 | 0.4448 | 0.7658 | 0.9467 |
|  | Epi | TW+XY | NJ | 0.8708 | 10.4297 | 0.4118 | 0.9370 | 0.8016 |

As shown in **Table 2**, FedDA-TSformer achieves superior performance on endocardium segmentation compared to TimeSformer, both with and without FedAvg. For the epicardium, FedDA-TSformer performance is comparable to other methods, even surpassing them when trained on the TW and XY datasets with NJ as the test set.

### 3.4 Hyperparameter settings

To identify the optimal hyperparameter values, we evaluated the model using various combinations of alpha and beta values from the set [0.0001, 0.01, 1, 100, 10000]. For endocardium segmentation, the best results were achieved with alpha = 0.01 and beta = 100, while for epicardium segmentation, the optimal performance was obtained with alpha = 0.0001 and beta = 0.0001. The results for other combinations are presented in **Figure 9**.

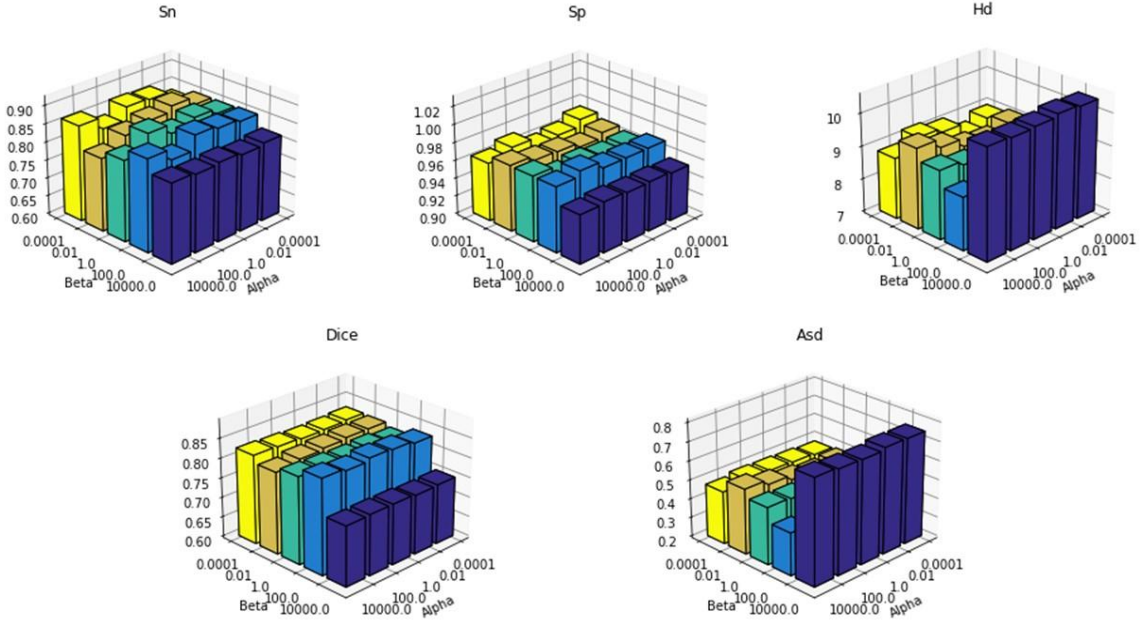

**Figure 9**. shows a 3D representation of the model's performance under different values of Alpha and Beta, which range from [0.0001, 0.01, 1, 100, 10000]. The horizontal axes correspond to the Alpha and Beta values, while the vertical axis shows the resulting performance metrics, including sensitivity (SN), specificity (SP), Hausdorff Distance (HD), Dice Similarity Coefficient (DSC), and Average Surface Distance (ASD). Each plot demonstrates how variations in these hyperparameters affect the performance of the model across different evaluation metrics.

We performed an ablation study to assess the impact of each component on the model, as shown in **Table 3**. The results indicate that the LMMD loss function has the most significant impact, particularly in enhancing boundary alignment metrics, including Hausdorff Distance (HD) and Average Surface Distance (ASD). The spatial self-attention mechanism alone does not achieve high segmentation accuracy, as measured by the Dice Similarity Coefficient (DSC), but performs significantly better when combined with other components. Furthermore, the temporal self-attention mechanism demonstrates optimal performance when integrated with the full set of modules. Overall, the complete model outperforms others across all evaluated metrics, except for specificity (SP), where it is slightly less effective.

**Table 3**. summarizes the results of an ablation study conducted to evaluate the impact of different model components: temporal attention (Time Att), spatial attention (Spatial Att), and Local Maximum Mean Discrepancy (LMMD) loss. The metrics presented include DSC (Dice Similarity Coefficient), HD (Hausdorff Distance in pixels), ASD (Average Surface Distance in pixels), SP (Specificity), and SN (Sensitivity). The arrows (↑ or ↓) indicate whether higher or lower values are preferred for each metric.

| Time Att | Spatial Att | LMMD | DSC↑ | HD(pixel) ↓ | ASD(pixel) ↓ | SP↑ | SN↑ |
|---|---|---|---|---|---|---|---|
|   |   |   | 0.8162 | 9.0560 | 0.5093 | 0.9700 | 0.8299 |
| x |   |   | 0.8263 | 9.0944 | 0.4793 | **0.9743** | 0.8247 |
|   | x |   | 0.8165 | 9.0530 | 0.5093 | 0.9614 | **0.8658** |
|   |   | x | 0.8349 | 8.7489 | 0.4460 | 0.9734 | 0.8398 |
| x | x |   | 0.8210 | 8.9020 | 0.4916 | 0.9657 | 0.8563 |
| x |   | x | 0.8352 | 8.5511 | 0.4398 | 0.9726 | 0.8430 |
|   | x | x | 0.8369 | 8.7476 | 0.4390 | 0.9741 | 0.8461 |
| x | x | x | **0.8423** | **8.3241** | **0.4089** | 0.9708 | 0.8647 |

We performed five-fold cross-validation for both endocardium and epicardium segmentation. Fold 0 consistently demonstrated the best performance for both, achieving high DSC, low HD and ASD, and favorable SP and SN metrics. The consistency of results across all folds indicates that the model is well-regularized and capable of achieving stable performance across different subsets of the data, as presented in **Table 4**.

**Table 4**. presents the results of a five-fold cross-validation for both endocardium (ENDO) and epicardium (EPI) segmentation. The metrics used include Dice Similarity Coefficient (DSC), Hausdorff Distance (HD in pixels), Average Surface Distance (ASD in pixels), Specificity (SP), and Sensitivity (SN). Fold 0 consistently demonstrates the best performance for both ENDO and EPI segmentation, achieving high DSC, low HD and ASD, as well as favorable SP and SN values. These results indicate that the model is well-regularized and capable of stable performance across different data subsets.

| Fold | Contour Type | DSC↑ | HD(pixel)↓ | ASD(pixel)↓ | SP↑ | SN↑ |
| --- | --- | --- | --- | --- | --- | --- |
| 0 | endocardium | 0.8423 | 8.3241 | 0.4089 | 0.9708 | 0.8647 |
| 1 | endocardium | 0.8396 | 8.8719 | 0.4334 | 0.9690 | 0.8629 |
| 2 | endocardium | 0.8162 | 8.8941 | 0.5444 | 0.9626 | 0.8585 |
| 3 | endocardium | 0.8364 | 9.3521 | 0.4656 | 0.9647 | 0.8544 |
| 4 | endocardium | 0.8196 | 9.3078 | 0.5328 | 0.9651 | 0.8516 |
| 0 | epicardium | 0.9078 | 8.5212 | 0.2545 | 0.8996 | 0.9176 |
| 1 | epicardium | 0.8396 | 8.8719 | 0.4334 | 0.8859 | 0.9165 |
| 2 | epicardium | 0.9018 | 9.1046 | 0.2770 | 0.8914 | 0.9068 |
| 3 | epicardium | 0.9014 | 9.5946 | 0.2977 | 0.8866 | 0.9124 |
| 4 | epicardium | 0.8973 | 9.6528 | 0.3113 | 0.8963 | 0.8913 |

To assess statistical significance, a t-test was conducted, with the null hypothesis assuming no significant difference in performance between the models, and the alternative hypothesis suggesting a significant difference. Statistical significance was evaluated for the proposed TimeSformer model against each baseline model using metrics such as DSC, HD, ASD, SN, and SP, as illustrated in Figures 10 and 11.

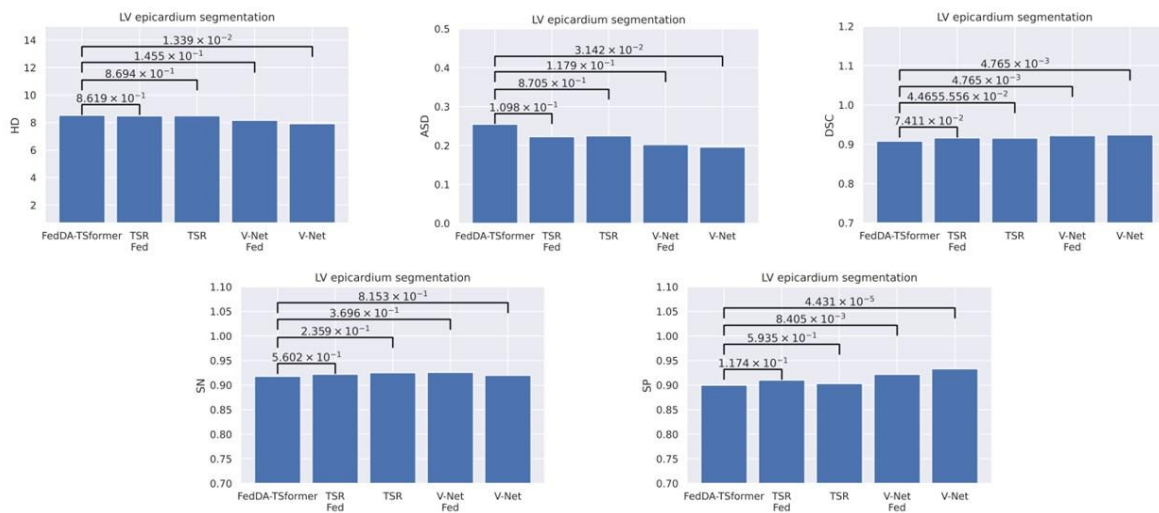

**Figure 10.** Illustration of the significant differences in performance for left ventricle epicardium segmentation among the baseline models across various evaluation metrics. Note that TSR represents TimeSformer. The horizontal axis represents the model names, while the vertical axis represents the performance values. For HD and ASD, lower values indicate better performance, whereas for DSC, SN, and SP, higher values are preferred.

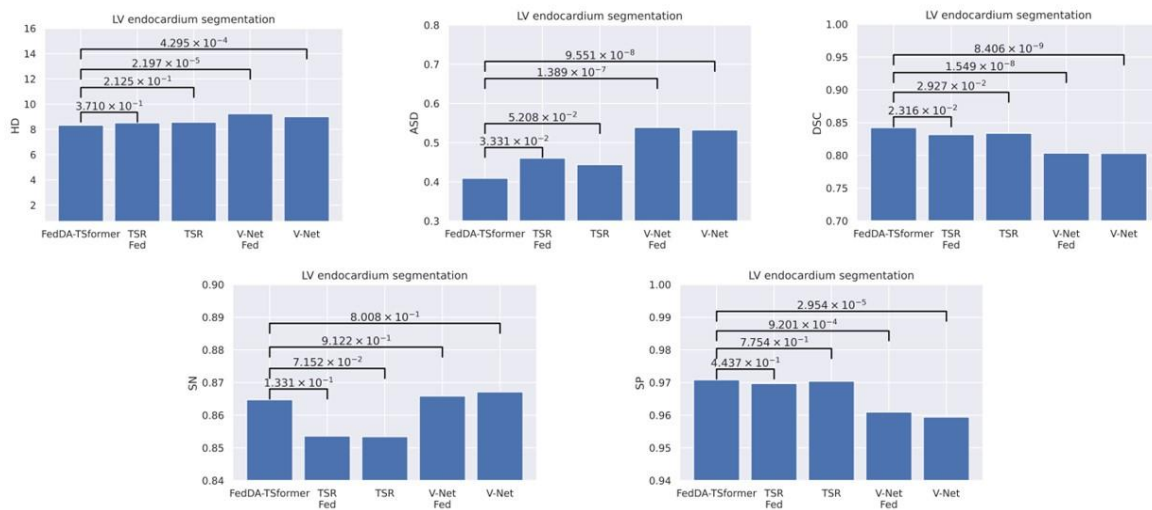

**Figure 11.** Illustration of the significant differences in performance for left ventricle endocardium segmentation among the baseline models across various evaluation metrics. Note that TSR represents TimeSformer. The horizontal axis represents the model names, while the vertical axis represents the performance values. For HD and ASD, lower values indicate better performance, whereas for DSC, SN, and SP, higher values are preferred.

## 4. Conclusion

We propose a modified TimeSformer capable of extracting left ventricular epicardial and endocardial contours while simultaneously updating the model using data from multiple sites through federated learning, without exchanging any original patient data. By leveraging TimeSformer's sensitivity to temporal and spatial data, our deep learning model achieves improved performance. This approach was validated on a dataset of 150 subjects, with several tests demonstrating its feasibility and accuracy. The proposed model ensures patient data confidentiality where privacy is required while effectively utilizing the data to provide auxiliary support for left ventricular endocardium and epicardium segmentation.

**Acknowledgement**:

This research was supported by an Interdisciplinary Seed Grants from Kennesaw State University (000149) and an AIREA grant from American Heart Association (25AIREA1377168).

**Declaration of Competing Interest**

The authors declare that they have no known competing financial interests or personal relationships that could have appeared to influence the work reported in this paper.

**Credit authorship contribution statement**

Yehong Huang: Conceptualization, methodology, coding, manuscript writing.

Chen Zhao: Supervision, project administration, funding acquisition, manuscript writing and review, approval.

Min Zhao, Guang-Uei Hung, and Zhixin Jiang: Data collection, manuscript writing, and review.

Weihua Zhou: Conceptualization, manuscript writing and review.